\documentclass[12pt]{article}
\usepackage{epsfig}
\usepackage{amssymb}
\usepackage{amsmath}
\usepackage{amsfonts}

\newcommand{\be}{\begin{equation}}
\newcommand{\ee}{\end{equation}}
\newcommand{\bea}{\begin{eqnarray}}
\newcommand{\eea}{\end{eqnarray}}

\newcommand{\kt}{\rangle}
\newcommand{\br}{\langle}

\newcommand{\ed}{\end{document}}

\newcommand{\np}{\newpage}

\newcommand{\bi}{\begin{itemize}}
\newcommand{\ei}{\end{itemize}}

\begin{document}

\title{Metric Operators for Quasi-Hermitian Hamiltonians and
Symmetries of Equivalent Hermitian Hamiltonians}
\author{\\
Ali Mostafazadeh
\\
\\
Department of Mathematics, Ko\c{c} University,\\
34450 Sariyer, Istanbul, Turkey\\ amostafazadeh@ku.edu.tr}
\date{ }
\maketitle

\begin{abstract}

We give a simple proof of the fact that every diagonalizable
operator that has a real spectrum is quasi-Hermitian and show how
the metric operators associated with a quasi-Hermitian Hamiltonian
are related to the symmetry generators of an equivalent Hermitian
Hamiltonian. \vspace{5mm}

\vspace{2mm}


\end{abstract}


\section{Introduction}

Given a separable Hilbert space ${\cal H}$ and a linear operator
$H:{\cal H}\to{\cal H}$ that has a real spectrum and a complete
set of eigenvectors, one can construct a new (physical) Hilbert
space ${\cal H}_{\rm phys}$ in which $H$ acts as a self-adjoint
operator. This allows for the formulation of a consistent quantum
theory where the observables and in particular Hamiltonian need
not be self-adjoint with respect to the standard ($L^2$-) inner
product on ${\cal H}$, \cite{jpa-2004b}. The physical Hilbert
space ${\cal H}_{\rm phys}$ and the observables are determined in
terms of a (bounded, everywhere-defined, invertible)
positive-definite metric operator $\eta_+:{\cal H}\to{\cal H}$
that renders $H$ pseudo-Hermitian \cite{p1}, i.e., $H$
satisfies\footnote{Here and throughout this article, we use
$A^\dagger$ to denote the adjoint of a linear operator $A$ that is
defined using the inner product $\br\cdot|\cdot\kt$ of ${\cal H}$
according to: $\br\psi|A\phi\kt=\br A\psi|\phi\kt$ for all
$\psi,\phi\in{\cal H}$.}
    \be
    H^\dagger=\eta_+ H\eta_+^{-1}.
    \label{ph}
    \ee
This marks the basic significance of the metric operator $\eta_+$.
The positivity of $\eta_+$ implies that $H$ belongs to a special
class of pseudo-Hermitian operators called quasi-Hermitian
operators \cite{quasi}.

The fact that for a given linear operator $H$ with a real
(discrete) spectrum and a complete set of eigenvectors, one can
always find a (positive-definite) metric operator $\eta_+$
fulfilling (\ref{ph}) has been established in \cite{p2} and the
role of antilinear symmetries such as ${\cal PT}$ has been
elucidated in \cite{p3}.\footnote{The alternative approach using
the so-called ${\cal CPT}$-inner product \cite{bender-prl-2002} is
equivalent to a specific choice of the metric operator
\cite{jmp-2003,jpa-2005a}.} Further investigation into the
properties of $\eta_+$ has revealed its non-uniqueness
\cite{npb-2002,jmp-2003,quasi} and the unitary-equivalence of $H$
and the Hermitian Hamiltonian
    \be
    h:=\rho\, H\rho^{-1},
    \label{h=111}
    \ee
where $\rho:=\sqrt\eta_+$, \cite{jpa-2003}.~\footnote{Given a
positive operator $X:{\cal H}\to{\cal H}$, $\sqrt X$ denotes the
unique positive square root of $X$.}  The latter observation has
been instrumental in providing an objective assessment of the
``complex (${\cal PT}$-symmetric) extension of quantum mechanics''
\cite{critique,fring-jpa-2006}. It has also played a central role
in clarifying the mysteries associated with the wrong-sign quartic
potential \cite{jones-mateo}. In short, the pseudo-Hermitian
quantum theory that is defined by the Hilbert space ${\cal H}_{\rm
phys}$ and the Hamiltonian $H$ admits an equivalent Hermitian
description in terms of the (standard) Hilbert space ${\cal H}$
and the Hermitian Hamiltonian $h$. However, the specific form of
$h$ depends on the choice of $\eta_+$. This has motivated the
search for alternative methods of computing the most general
metric operator for a given $H$,
\cite{jpa-2006b,scholtz-geyer-plb,jmp-2006a,fring-cjp-2006,tjp-2006,musumbu-jpa-2007}.

In this article we first give a simple proof of the existence of
metric operators $\eta_+$ and then relate $\eta_+$ to the
symmetries of an equivalent Hermitian Hamiltonian.

\section{Existence of Metric Operators}

Let $H:{\cal H}\to{\cal H}$ be a (closed) operator with a real
spectrum, and suppose that it is diagonalizable, i.e., there are
operators $T,H_d:{\cal H}\to{\cal H}$ such that $T$ is invertible
(bounded and hence closed),
    \be
    H=T^{-1}H_d\, T,
    \label{diag}
    \ee
and $H_d$ is diagonal in some orthonormal basis of ${\cal H}$. The
latter property implies that $H_d$ is a normal operator.
Furthermore, because $H$ and $H_d$ are isospectral, the spectrum
of $H_d$ is also real. This together with the fact that $H_d$ is
normal imply that it is Hermitian (self-adjoint).

Next, recall that because $T$ is a closed, invertible operator it
admits a polar decomposition \cite{reed-simon}:
    \be
    T=U\,\rho,
    \label{polar}
    \ee
where $U$ is a unitary operator and $\rho=|T|:=\sqrt{T^\dagger T}$
is invertible and positive(-definite). Inserting, (\ref{polar})
into (\ref{diag}) and introducing
    \be
    h:=U^\dagger H_d U,
    \label{h=}
    \ee
we find
    \be
    H=\rho^{-1}h\rho.
    \label{H=H}
    \ee
Because $\rho$ is positive-definite, so is $\eta_+:=\rho^2$.
Because $H_d$ is Hermitian and $U$ is unitary, $h$ is Hermitian.
In view of this and the fact that $\rho$ is also Hermitian,
(\ref{H=H}) implies $H^\dagger=\eta_+H\eta_+^{-1}$. This proves
the existence of a metric operator $\eta_+$ that makes $H$,
$\eta_+$-pseudo-Hermitian.

The above proof is shorter than the one given in \cite{p2}. But it
has the disadvantage that it does not offer a method of computing
$\eta_+$.

\section{Metric Operators and Symmetry Generators}

Let $\eta_+$ and $\eta_+'$ be a pair of metric operators rendering
$H$ pseudo-Hermitian, $\rho:=\sqrt\eta_+$, and
$\rho':=\sqrt{\eta_+'}$. Then the Hermitian Hamiltonian operators
    \be
    h:=\rho H\rho^{-1},~~~~~~~~h':=\rho' H{\rho'}^{-1}
    \label{hh}
    \ee
are unitary-equivalent to $H$, \cite{jpa-2003}. It is easy to see
that $h$ and $h'$ are related by the similarity transformation
    \be
    h'=A\, h\, A^{-1},
    \label{sim}
    \ee
where
    \be
    A:=\rho'\rho^{-1}.
    \label{A=}
    \ee
Now, taking the adjoint of both sides of (\ref{sim}) and using the
fact that $h$ and $h'$ are Hermitian, we find
    \be
    [A^\dagger A,h]=0.
    \label{sym}
    \ee
This means that $A^\dagger A$ is a (positive-definite) symmetry
generator for the Hamiltonian $h$. Furthermore, (\ref{A=}) and
$\eta'_+={\rho'}^{2}$ lead to the curious relation:
    \be
    \eta'_+=\rho\: A^\dagger A\:\rho.
    \label{eta-prime}
    \ee
Another immediate consequence of (\ref{A=}) is
    \be
    A^\dagger=\rho^{-1}A\,\rho,
    \label{A-ph}
    \ee
i.e., $A$ is $\rho^{-1}$-pseudo-Hermitian \cite{p1}.

It is easy to show that the converse relationship also holds,
i.e., given an invertible linear operator $A:{\cal H}\to {\cal H}$
that satisfies (\ref{A=}) and (\ref{A-ph}), the operator $\eta'$
defined by
    \be
    \eta'_+:=\rho\: A^\dagger A\:\rho.
    \label{eta-def}
    \ee
renders $H$, $\eta'$-pseudo-Hermitian.

The above analysis shows that given a metric operator
$\eta_+=\rho^2$ for the Hamiltonian $H$, we can express any other
metric operator for $H$ in the form
    \be
    \eta'_+=\rho\: S\:\rho,
    \label{eta-form}
    \ee
where $S$ is a positive-definite symmetry generator of $h$ such
that there is a $\rho^{-1}$-pseudo-Hermitian operator $A$
satisfying
    \be
    S=A^\dagger A.
    \label{S}
    \ee

In practice, the construction of the symmetry generators $S$ of
the Hermitian operator $h$ is easier than that of the
$\rho^{-1}$-pseudo-Hermitian operators $A$. This calls for a
closer look at the structure of $A$.

In view of (\ref{S}), we can express $A$ in the form
    \be
    A=U\:\sigma,
    \label{A=US}
    \ee
where $U:{\cal H}\to{\cal H}$ is a unitary operator and
$\sigma:=\sqrt S$. This reduces the characterization of $A$ to
that of appropriate unitary operators $U$ that ensure
$\rho^{-1}$-pseudo-Hermiticity of $A$.

Inserting (\ref{A=US}) in (\ref{A-ph}) and introducing
    \be
    B:=\rho\:U,
    \label{B}
    \ee
we find
    \be
    B^\dagger=\sigma\:B\sigma^{-1}.
    \label{B-ph}
    \ee
That is, $B$ is $\sigma$-pseudo-Hermitian. Moreover, (\ref{B})
implies
    \be
    \eta_+=BB^\dagger.
    \label{eta=BB}
    \ee
Conversely, given a positive-definite symmetry generator $S$ and a
$\sqrt S$-pseudo-Hermitian operator $B$ satisfying (\ref{eta=BB}),
we can easily show that the operator
    \be
    U:=\rho^{-1}B,
    \label{U=}
    \ee
is unitary and $A$ given by (\ref{A=US}) is
$\rho^{-1}$-pseudo-Hermitian. As a result, the most general metric
operator $\eta_+'$ is given by (\ref{eta-form}), alternatively
    \be
    \eta_+'=(\sqrt S\,\rho)^\dagger(\sqrt S\,\rho),
    \label{eta-prime-3}
    \ee
where $S$ is a positive-definite symmetry generator of $h$ such
that there is a $\sqrt S$-pseudo-Hermitian operator $B$ satisfying
$\eta_+=BB^\dagger$.

\np

\section{Concluding Remarks}

The existence of a positive-definite metric operator $\eta_+$ that
renders a diagonalizable Hamiltonian operator $H$ with a real
spectrum $\eta_+$-pseudo-Hermitian can be directly established
using the well-known polar decomposition of closed operators.
Previously, we have pointed out that one can describe the most
general $\eta_+$ in terms of a given metric operator and certain
symmetry generators $A$ of $H$, \cite{jmp-2003}. Here we offer
another description of the most general $\eta_+$ in terms of
certain positive-definite symmetry generators $S$ of a given
equivalent Hamiltonian $h$. Unlike the symmetry generators $A$ of
$H$ that are non-Hermitian, the operators $S$ are standard
Hermitian symmetry generators. This makes the results of this
paper more appealing.

For the cases that $h$ is an element of a dynamical Lie algebra
${\cal G}$ with ${\cal H}$ furnishing a unitary irreducible
representation of ${\cal G}$, one can identify the
positive-definite symmetry generators $S$ with certain functions
of a set of mutually commuting elements of ${\cal G}$ that
includes $h$. For example, one can construct $S$ for the two-level
system, where ${\cal G}=u(2)$, or the generalized (and simple)
Harmonic oscillator where ${\cal G}=su(1,1)$, \cite{nova}. These
respectively correspond, to the general two-level quasi-Hermitian
Hamiltonians \cite{tjp-2006} and the class of quasi-Hermitian
Hamiltonians that are linear combinations of $x^2$, $p^2$, and
$\{x,p\}$ such as the one considered in \cite{swanson}. For these
systems one can also construct a metric operator $\eta_+$ and its
positive square root $\rho$. Nevertheless the implementation of
the formula (\ref{eta-form}) for obtaining the most general metric
operator proves impractical. This is because it is not easy to
characterize the general form of $\sqrt S$-pseudo-Hermitian
operators $B$ that would fulfil $\eta_+=BB^\dagger$.

Although the formula (\ref{eta-form}) seems to be of limited
practical value, it is conceptually appealing because it traces
the non-uniqueness of the metric operator to the symmetries of the
equivalent Hermitian Hamiltonians.

\np

\ed